%% file: main.tex
\documentclass[letterpaper]{article} 
\usepackage{aaai2026}  
\usepackage{times}  
\usepackage{helvet}  
\usepackage{courier}  
\usepackage[hyphens]{url}  
\usepackage{graphicx} 
\usepackage{natbib}  
\usepackage{caption} 
\usepackage{algorithm}
\usepackage{algorithmic}

\usepackage{amsmath} 
\usepackage{amsfonts} 
\usepackage{booktabs}
\usepackage{multirow} 
\usepackage{newfloat}
\usepackage{listings}
\DeclareCaptionStyle{ruled}{labelfont=normalfont,labelsep=colon,strut=off} 
\floatstyle{ruled}
\newfloat{listing}{tb}{lst}{}
\floatname{listing}{Listing}
\title{BAMAS: Structuring Budget-Aware Multi-Agent Systems}
\author{
    Liming Yang\textsuperscript{\rm 1},
    Junyu Luo\textsuperscript{\rm 1},
    Xuanzhe Liu\textsuperscript{\rm 1},
    Yiling Lou\textsuperscript{\rm 2},
    Zhenpeng Chen\textsuperscript{\rm 3,4}\thanks{Corresponding author: Zhenpeng Chen.}
}
\affiliations{
    \textsuperscript{\rm 1}Peking University, 
    \textsuperscript{\rm 2}University of Illinois Urbana-Champaign,
    \textsuperscript{\rm 3}Nanyang Technological University,
    \textsuperscript{\rm 4}Tsinghua University
    \\
    yangliming2018@pku.edu.cn, luojunyu@stu.pku.edu.cn, liuxuanzhe@pku.edu.cn, yilingl@illinois.edu, zhenpeng.chen@ntu.edu.sg
}

\usepackage{bibentry}
\usepackage{xcolor}

\begin{document}

\maketitle

\input{1_abs}

\input{1.5Introduction}
\input{2_related}
\input{3_method}

\input{4_experiments}
\input{5_conclusion}

\section{Acknowledgement}
This work was supported by the National Natural Science Foundation of China under the grant number 62325201.


\bibliography{aaai2026}
\input{appendix}

\end{document}

%% file: 1_abs.tex
\begin{abstract}
Large language model (LLM)-based multi-agent systems have emerged as a powerful paradigm for enabling autonomous agents to solve complex tasks. As these systems scale in complexity, cost becomes an important consideration for practical deployment. However, existing work rarely addresses how to structure multi-agent systems under explicit budget constraints. In this paper, we propose BAMAS, a novel approach for building multi-agent systems with budget awareness. BAMAS first selects an optimal set of LLMs by formulating and solving an Integer Linear Programming problem that balances performance and cost. It then determines how these LLMs should collaborate by leveraging a reinforcement learning-based method to select the interaction topology. Finally, the system is instantiated and executed based on the selected agents and their collaboration topology. We evaluate BAMAS on three representative tasks and compare it with state-of-the-art agent construction methods. Results show that BAMAS achieves comparable performance while reducing cost by up to 86\%.

\end{abstract}

%% file: 1.5Introduction.tex
\section{Introduction}
LLM-based multi-agent systems, in which LLMs serve as core reasoning engines that perceive, communicate, and collaborate, have demonstrated remarkable capabilities in solving complex tasks~\cite{wu2023autogen, hong2023metagpt}. By supporting diverse interaction patterns, these systems can decompose problems, generate alternative solutions, and iteratively refine outputs, thereby pushing the boundaries of automated problem solving.

However, as these systems increase in complexity and agent count, cost becomes a critical consideration for practical deployment~\cite{chen2025surveyllmbasedmultiagentsystem}.  In particular, the cost is primarily driven by token consumption during LLM calls. A single task may require dozens of LLM calls across multiple agents, with costs scaling unpredictably depending on the collaboration topology and reasoning depth. Such unpredictability in cost makes these systems difficult to scale reliably in production.

Despite the critical role of budget management, existing research has predominantly focused on maximizing performance, often treating cost as an afterthought. Leading frameworks such as AutoGen~\citep{wu2023autogen} and MetaGPT~\citep{hong2023metagpt} typically employ reactive strategies that provide limited control over the cost-performance trade-off. To our knowledge, no prior work directly addresses the fundamental question: how can we design a multi-agent system that delivers strong task performance while adhering to a predefined cost budget?

To address this challenge, we propose \textbf{BAMAS}, a novel approach for constructing budget-aware multi-agent systems. Our key insight is that since LLMs are the primary drivers of cost, effective budget management should start by treating LLM allocation as a constrained optimization problem. Accordingly, BAMAS begins with selecting an optimal set of LLMs by formulating and solving an Integer Linear Programming problem that ensures both performance and strict adherence to the given budget. Next, BAMAS learns how to orchestrate collaboration among the selected LLMs by training a topology selection policy via reinforcement learning. This policy identifies the most effective collaboration topology, such as linear or star structures, that maximizes task performance within the allocated resources. Finally, BAMAS instantiates the multi-agent system based on the chosen LLM pool and collaboration topology, strategically balancing the cost-performance trade-off.

We evaluate BAMAS on three widely-used benchmark datasets spanning code generation and mathematical reasoning, and compare it against three state-of-the-art multi-agent construction approaches. Results show that BAMAS achieves comparable accuracy while reducing costs by up to 86\% and adhering to predefined budgets, demonstrating a strong cost-performance trade-off. An ablation study further confirms that our joint optimization strategy significantly outperforms greedy baselines by identifying superior solutions that balance cost and performance. Our analysis also reveals that BAMAS learns to adaptively select collaboration topologies: it favors simpler topologies under tight budgets, while tailoring its structure to specific task domains when resources are more ample, offering both adaptability and interpretability.

In summary, this paper makes the following contributions:
\begin{itemize}
\item We introduce BAMAS, a novel framework for constructing multi-agent systems under budget constraints. BAMAS jointly optimizes LLM selection and agent collaboration topology through Integer Linear Programming and reinforcement learning, respectively, to maximize task performance within a fixed cost budget.
\item We evaluate BAMAS on three widely-used datasets and compare it against three state-of-the-art agent construction approaches. BAMAS achieves comparable performance while reducing overall cost by up to 86\%, demonstrating an effective cost-performance trade-off.
\item We publicly release our code and data at \url{https://github.com/chunfenri/BAMAS}, to support further research.
\end{itemize}

%% file: 2_related.tex
\section{Related Work}
\noindent \textbf{Multi-agent systems.} LLM-based multi-agent systems
have emerged as a powerful paradigm for enabling autonomous agents to tackle complex tasks. These systems have gained increasing attention from both academia and industry, prompting the development of frameworks that investigate how to structure and coordinate multiple agents effectively~\cite{zhou2024survey}. Notable examples include MetaGPT \cite{hong2023metagpt}, AutoGen \cite{wu2023autogen}, and ChatDev \cite{qian2024chatdevcommunicativeagentssoftware}. MetaGPT adopts a meta-programming paradigm to enforce standardized workflows, where agents with specialized roles follow predefined steps in a rigid execution pipeline. AutoGen offers a more flexible agent architecture, enabling the composition of multi-agent systems through custom dialogue loops, role definitions, and dynamic turn-taking protocols. ChatDev simulates a virtual company, with agents collaborating through structured conversations to complete end-to-end tasks. While these frameworks demonstrate strong performance on various problems, they pay limited attention to cost-efficiency, a key concern in real-world deployments. In contrast, we focus on budget-aware multi-agent structuring, aiming to achieve a favorable trade-off between performance and cost.

\noindent \textbf{Cost-efficiency AI.} The rising computational and monetary costs of modern AI systems have spurred research on cost-efficiency, focusing on optimizing performance under explicit cost or budget constraints \cite{chen2023frugalgpt,zhang2024treacle,dekoninck2025cascade}. Budget-aware computing has thus emerged as a field that treats cost consumption as a primary factor in system design and optimization \cite{zhang2024treacle,dekoninck2025cascade}. Various advanced methods have been developed to effectively navigate the trade-offs between performance and incurred costs \cite{li2025budgetguidance,zellinger2025earlyabstention,arora2025efficientreasoning}. As a rapidly evolving paradigm, LLM-based multi-agent systems also face such cost challenges. This paper takes an important step towards addressing this gap by proposing a budget-aware multi-agent structuring approach that balances task performance with token consumption costs.

%% file: 3_method.tex
\begin{figure*}[t]
    \centering
    \includegraphics[width=\textwidth]{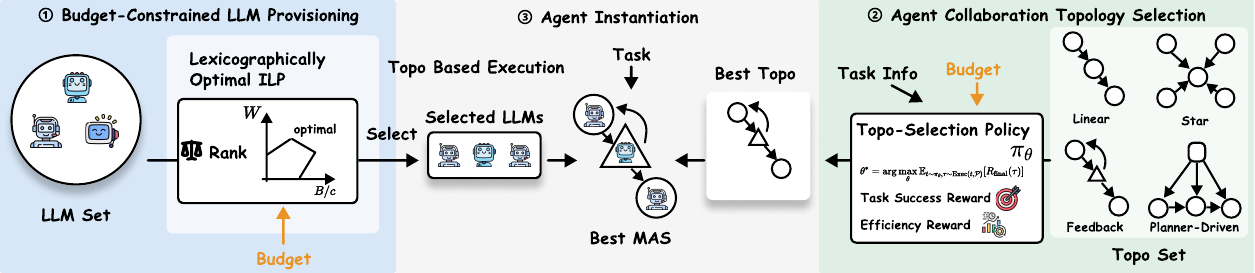}
    \caption{Overview of BAMAS, which constructs a budget-aware multi-agent system by provisioning a cost-optimal set of LLMs and selecting the best collaboration topology to guide task execution. 
    }
    \label{fig:framework}
\end{figure*}

\section{Methodology}
We formulate the core problem as follows: given a task $T$, a set of available LLMs $\mathcal{A}$, and a cost budget $B$, the goal is to structure a multi-agent system that achieves a favorable performance–cost trade-off under the budget constraint. Specifically, we aim to determine an optimal subset of LLMs $\mathcal{P} \subseteq \mathcal{A}$ and a collaboration topology $t$ that governs their interactions. Each selected LLM in $\mathcal{P}$ functions as a distinct agent in the system.

To address this problem, we propose \textbf{BAMAS}, a novel approach for structuring budget-aware multi-agent systems. As illustrated in Figure~\ref{fig:framework}, BAMAS consists of three key components:  
(1)~\emph{budget-constrained LLM provisioning}, which formulates the selection of an optimal LLM pool \( \mathcal{P} \) as an Integer Linear Programming (ILP) problem and solves it to satisfy the given cost budget;  
(2)~\emph{agent collaboration topology selection}, which designs a reinforcement learning-based method to identify an effective agent interaction topology \( t \); and  
(3)~\emph{agent instantiation}, where the multi-agent system is instantiated and orchestrated based on the selected \( \mathcal{P} \) and \( t \) to perform the task.

\subsection{Budget-Constrained LLM Provisioning}
This component selects a subset of LLMs $\mathcal{P}$ from the available pool $\mathcal{A}$ such that the total cost does not exceed a given budget $B$. Empirical studies have shown that, for complex reasoning tasks, a single high-performance model often outperforms ensembles of weaker models~\cite{wang-etal-2024-rethinking-bounds}. Motivated by this insight, we adopt a performance-first selection strategy that prioritizes stronger LLMs whenever they can be afforded within the budget.

We formalize the selection as an ILP problem, aiming to maximize overall performance within the cost budget.

To operationalize LLM performance, we rank all available LLMs (note that multiple instances of the same LLM may exist) into tiers, with \( \mathcal{A}_1 \) representing the highest-performance tier and higher indices indicating lower tiers, down to \( \mathcal{A}_L \) as the lowest-performance tier. We use the widely referenced LMSys Chatbot Arena Leaderboard~\cite{chiang2024chatbotarenaopenplatform} as a performance proxy, where a higher leaderboard rank indicates better performance.

The cost of the $i$-th LLM \( \mathcal{A}_i \) is computed as: 

\begin{equation}
c_i = T_{in} \cdot P_{in} + T_{out} \cdot P_{out},
\end{equation}

where $T_{\text{in}}$ and $T_{\text{out}}$ denote the input and output token counts per API call for task $T$, and $P_{\text{in}}$, $P_{\text{out}}$ are the corresponding per-token prices provided by the LLM vendor. Since exact token usage is unknown before execution, we adopt representative estimates.
Specifically, we set \( T_{\text{in}} = 500 \), based on prior measurements showing that typical LLM input lengths fall within 128$\sim$256 tokens~\cite{li2025largelanguagemodelinference}, and we roughly double the upper bound (256 $\rightarrow$ 500) to accommodate additional context an agent may ingest from others in multi-agent settings; we determine \( T_{\text{out}} \) by sampling 50 training instances from the target dataset and using the maximum observed output length as an upper bound.

We then construct a decision weight \( W_i \) for each LLM tier to guide the ILP solver in prioritizing higher-tier LLMs. These weights are defined recursively in a bottom-up manner to ensure that selecting a higher-tier LLM always outweighs any budget-feasible combination of lower-tier LLMs. Specifically, we set the weight for the lowest tier as \( W_L = 1 \), and compute the weight for any higher tier \( i \) as:

\begin{equation}
W_i = 1 + \sum_{j=i+1}^{L} \left( W_j \cdot \left\lfloor \frac{B}{c_j} \right\rfloor \right).
\end{equation}

This construction guarantees that any single LLM from tier \( \mathcal{A}_i \) has a higher total weight than any group of lower-tier LLMs that can be afforded under the budget \( B \).

Let \( x_{ij} \) be a binary decision variable, where \( x_{ij} = 1 \) indicates selecting the \( j \)-th instance of the \( i \)-th tier LLM \( \mathcal{A}_i \). This allows that multiple instances of the same LLM are selected.  
The objective is to maximize the total weight of the selected LLMs, subject to the budget constraint $B$ and configuration requirement—specifically, to ensure a meaningful multi-agent setup, at least two LLMs must be selected.
The resulting ILP formulation is:

\begin{equation}
\label{eq:ilp}
\begin{aligned}
& \underset{x}{\text{maximize}}
& & \sum_{i=1}^{L} \sum_{j} W_i \cdot x_{ij} ,\\
& \text{subject to}
& & \sum_{i=1}^{L} \sum_{j} c_i \cdot x_{ij} \leq B,
& \sum_{i=1}^{L} \sum_{j} x_{ij} \geq 2.
\end{aligned}
\end{equation}

Solving this ILP yields the budget-feasible candidate LLM pool \( \mathcal{P} \). Due to the structure of \( W_i \), the selected configuration is guaranteed to be lexicographically optimal, maximizing overall performance within the budget. This forms a strong foundation for the downstream components, ensuring that subsequent agent construction operates on the best-available LLM composition.

\subsection{Agent Collaboration Topology Selection}
While the provisioning component determines \textit{which} LLMs to use, this component decides \textit{how} these LLMs should collaborate to solve the given task. A fixed collaboration topology is unlikely to perform well across a wide variety of tasks. To address this, we introduce a topology selection policy, \( \pi_\theta \), that selects an appropriate agent collaboration topology \( t \in \mathcal{T} \), conditioned on the task specification \( T \) and the total budget \( B \).

The action space of \( \pi_\theta \) is a curated library \( \mathcal{T} \) of widely-used agent workflow topologies, each reflecting a distinct cognitive or reasoning strategy:

\begin{itemize}
    \item \textbf{Linear Topology:} Implements sequential reasoning, where each agent builds upon the output of the previous one. This topology is well-suited for multi-step reasoning tasks~\cite{hong2023metagpt, qian2024chatdevcommunicativeagentssoftware}.
    
    \item \textbf{Star Topology:} Supports parallel hypothesis generation and evaluation, embodying a divide-and-conquer approach ideal for decomposable or multi-perspective problems~\cite{du2023mad,zheng2023judge}.
    
    \item \textbf{Feedback Topology:} Enables iterative refinement through generate-and-critique cycles. This is particularly effective for tasks requiring self-correction or quality enhancement~\cite{madaan2023selfrefine,shinn2023reflexion}.
    
    \item \textbf{Planner-Driven Topology:} Incorporates a central planner agent to dynamically coordinate other agents, offering maximum flexibility. It is especially useful for open-ended or unstructured tasks with no predefined solution path~\citep{yao2022react,yao2023treeofthoughts}.
\end{itemize}

We train the topology selection policy \( \pi_\theta \) using an offline reinforcement learning paradigm. This design choice is driven by the prohibitive cost and latency associated with collecting online trajectories in LLM-based multi-agent environments. The learning objective is to optimize the policy parameters \( \theta \) to maximize the expected final reward derived from executing a complete task trajectory \( \tau \). Formally, the objective is given by:
\begin{equation}
\begin{split}
\theta^* & = \arg\max_{\theta} J(\theta) \\
& = \arg\max_{\theta} \mathbb{E}_{t \sim \pi_\theta, \tau \sim \text{Exec}(t, \mathcal{P})} \left[ R_{\text{final}}(\tau) \right].
\end{split}
\end{equation}
Here, \texttt{Exec}$(t, \mathcal{P})$ denotes the agent execution engine that instantiates and runs the chosen topology \( t \) using the previously provisioned LLM pool \( \mathcal{P} \), and \( R_{\text{final}}(\tau) \) is a composite reward function reflecting both task success and cost efficiency.
In the following, we describe the composite reward function and the learning process in detail.

\subsubsection{Composite Reward Function}
The final reward \( R_{\text{final}}(\tau) \) is designed as a weighted combination of two distinct components. The weights \( (w_{\text{perf}}, w_{\text{cost}}) \) are tunable hyperparameters that govern the trade-off between maximizing task performance and minimizing computational cost:
\begin{equation}
R_{\text{final}}(\tau) = w_{\text{perf}} \cdot R_{\text{perf}} + w_{\text{cost}} \cdot R_{\text{cost}}.
\end{equation}

Each reward component serves a different yet complementary role:

\begin{itemize}
    \item \textbf{Task Success Reward (\( R_{\text{perf}} \)):} This is a sparse, binary reward that evaluates whether the task was solved correctly. It provides a clear learning signal for achieving the primary goal: \( R_{\text{perf}} = +C_{\text{succ}} \) if the task succeeds, and \( -C_{\text{fail}} \) otherwise.

    \item \textbf{Cost Efficiency Reward (\( R_{\text{cost}} \)):} This component promotes budget adherence and cost-efficient solutions. A heavy penalty \( -C_{\text{overflow}} \) is incurred if the actual cost \( C_{\text{actual}}(\tau) \) exceeds the budget \( B \). If the task is successful, an additional bonus \( g(1 - C_{\text{actual}}(\tau)/B) \) is granted, which scales with the amount of budget saved. Importantly, this efficiency bonus is awarded \emph{only} for successful executions. This encourages the policy to find solutions that are not only correct but also efficient, while discouraging degenerate strategies that reduce cost at the expense of task performance. 
\end{itemize}

\subsubsection{Learning Algorithm and Loss Function}
Our learning algorithm is a form of REINFORCE \cite{williams1992reinforce}, tailored for an offline, one-shot decision context where the policy makes a single choice of topology, $t$, and receives a final reward, $R_{\text{final}}$, after the entire trajectory completes. We train the policy by minimizing the following loss function on a static dataset, $\mathcal{D}$, of pre-collected experiences:

\begin{equation}
\begin{split}
\mathcal{L}(\theta) &= - \hat{\mathbb{E}}_{(T, B, t, \tau) \sim \mathcal{D}} \left[ \log \pi_\theta(t|T, B) \cdot R_{\text{final}}(\tau) \right] \\ 
& \quad - \beta \cdot H(\pi_\theta(\cdot|T, B)).
\end{split}
\label{eq:loss}
\end{equation}

This loss function consists of two key terms that guide the optimization:
\begin{enumerate}
    \item \textbf{Policy Gradient Term:} The term $-\log \pi_\theta(t|T, B) \cdot R_{\text{final}}(\tau)$ is the core of the learning algorithm. Here, the final trajectory reward, $R_{\text{final}}(\tau)$, serves as an estimate of the advantage of selecting topology $t$. By minimizing this negative log-likelihood weighted by the reward, the optimization process updates $\theta$ to increase the probability of actions that lead to high rewards. This approach directly connects the policy's decision to its ultimate outcome, providing a clear and powerful learning signal without the high variance often introduced by importance sampling in offline settings \citep{thomas2016data,xie2019mis}.
    
    \item \textbf{Entropy Regularization Term:} The term $-\beta \cdot H(\pi_\theta(\cdot|T, B))$ is an entropy bonus, where $H(\cdot)$ is the Shannon entropy and $\beta$ is a hyperparameter. This term encourages the policy to maintain a degree of stochasticity, preventing it from prematurely converging to a single, suboptimal topology. By promoting exploration within the policy space, it enhances training stability and improves the chances of discovering a more globally optimal strategy. 
\end{enumerate}

Given the reward function, learning algorithm, and loss function, we perform training of the topology selection policy.
During training, the model parameters $\theta^*$ that achieve the highest average composite reward across the entire offline dataset $\mathcal{D}$ are preserved as the final optimized policy. 
The complete training procedure is presented in Appendix~A.

\subsection{Agent Instantiation}
The final component constructs the multi-agent system by instantiating the selected LLMs \(\mathcal{P}\) within the chosen workflow topology \(t \). Its primary function is to assign specific LLM instances from \(\mathcal{P}\) to the distinct agent roles defined by the topology. In Linear and Star, all agents act as executors. In Feedback, the highest-weight LLMs serve as critics and the rest as executors. In Planner-driven, the highest- and second-highest-weight LLMs serve as the planner and critics, respectively, and the remaining models act as executors. Since higher weight reflects stronger capability, this assignment ensures that the most capable LLMs are allocated to the most critical roles.

Once instantiated, the execution engine manages information flow and coordination among agents based on the selected topology. For Linear and Star, the engine uses a template-based scheduler to invoke agents sequentially, minimizing coordination overhead. For Feedback, the engine pairs a dedicated critic with one or more executors to run a generate–critique–revise loop: the critic only audits outputs and does not perform task-side computation; if no issues are found, the executor’s latest output is returned as the final answer, otherwise the critique is routed back for revision until acceptance or a preset budget limit. For Planner-driven, the planner dynamically orchestrates agent interactions step by step according to evolving requirements.

%% file: 4_experiments.tex
\newcommand{\finding}[1]{
    \vspace{2mm}
    \noindent
    \fbox{
        \begin{minipage}{0.97\linewidth}
        \textbf{Finding:} #1
        \end{minipage}
    }
    \vspace{2mm}
}
\section{Evaluation Setup}

\subsection{Research Questions (RQs)}
We evaluate BAMAS by answering the following RQs.

\noindent \textbf{RQ1 (Cost–performance trade-off):} How well does BAMAS balance cost and task performance compared to existing multi-agent system construction approaches?

\noindent \textbf{RQ2 (Component analysis):} How essential are the core components of BAMAS (i.e., LLM provisioning and topology selection) compared to a simplified and greedy cost-aware strategy?

\noindent \textbf{RQ3 (Topology selection):} Can BAMAS, through its topology selection policy, choose topologies in a way that adapts to both task requirements and budget constraints?

\subsection{Datasets}
We evaluate BAMAS on three widely used benchmarks, using each dataset to generate reinforcement learning training data and conduct evaluation.

\begin{itemize}
    \item \textbf{GSM8K}~\citep{cobbe2021trainingverifierssolvemath}: A dataset of grade school math word problems. We use the first 1,000 examples from the official training set as the reinforcement learning training corpus and evaluate on the full test set of 1,319 problems.

    \item \textbf{MBPP}~\citep{austin2021programsynthesislargelanguage}: A dataset for Python programming tasks. We use all 374 problems in the training set as the reinforcement learning training corpus and evaluate on the full 500-problem test set.

    \item \textbf{MATH}~\citep{hendrycks2021measuringmathematicalproblemsolving}: A challenging dataset of advanced mathematical reasoning problems, large and diverse in difficulty levels and problem types. To manage its size, we perform stratified sampling by difficulty level and problem type to construct training and test sets of 1,000 problems each, preserving the original distribution.
\end{itemize}

\subsection{Available LLMs}
LLMs are the foundational components of our multi-agent systems. In our experiments, we select two representative LLMs that illustrate the trade-off between performance and cost-efficiency:

\begin{itemize}
    \item \textbf{DeepSeek-V3:} A high-performance model priced at \$0.27 per million input tokens and \$1.10 per million output tokens.
    \item \textbf{GPT-4.1 nano:} A lower-performance but more cost-effective alternative priced at \$0.10 per million input tokens and \$0.40 per million output tokens.
\end{itemize}

\subsection{Evaluation Metrics}
To evaluate the effectiveness and efficiency of multi-agent systems, we adopt the following metrics:

\begin{itemize}
    \item \textbf{Accuracy (Acc \%):} The task success rate, defined as the percentage of tasks for which the constructed multi-agent system produces a correct solution.

    \item \textbf{Average Cost (Avg Cost):} The mean cost of the multi-agent system across all tasks. For each task, the cost is computed as \begin{equation}
        \text{Cost} = \left( \sum_{i \in \text{calls}} P_{\text{in},i} \cdot T_{\text{in},i} + \sum_{i \in \text{calls}} P_{\text{out},i} \cdot T_{\text{out},i} \right) \times 10^6,
    \end{equation}
    where each API call $i$ contributes cost based on its input and output token counts ($T_{\text{in},i}$ and $T_{\text{out},i}$) and corresponding per-token prices ($P_{\text{in},i}$ and $P_{\text{out},i}$). The factor $10^6$ is used to rescale the metric for clearer presentation.
\end{itemize}

\subsection{Baselines}
We compare BAMAS against three state-of-the-art multi-agent system construction approaches: \textbf{AutoGen}~\citep{wu2023autogen}, \textbf{MetaGPT}~\citep{hong2023metagpt}, and \textbf{ChatDev}~\citep{qian2024chatdevcommunicativeagentssoftware}. Detailed descriptions of these approaches are provided in the Related Work section. We use these baselines for answering RQ1.

Additionally, we introduce a heuristic baseline, \textbf{Naive-CostAware}, which greedily selects agent configurations according to five predefined resource levels (Levels 1–5). Each level corresponds to an increasing number of LLMs, with Level~$i$ using $(i+1)$ LLMs, as a multi-agent system requires at least two. Unlike BAMAS, this baseline performs no global optimization via ILP and does not conduct topology selection. We employ this baseline to address RQ2.

Since none of the baselines incorporates an LLM selection component like BAMAS, we fix the LLM type when employing these approaches. To ensure fair comparison, all baselines are restricted to use DeepSeek-V3 and GPT-4.1 nano, matching the LLMs used by BAMAS. For example, when applying AutoGen, we implement two versions (one with DeepSeek-V3 and another with GPT-4.1 nano) and include both in the comparison with BAMAS. For the MATH dataset, which contains advanced problems that incur substantially higher costs, we report baseline results using only GPT-4.1 nano to reduce expenses.

\begin{table}[t]
\centering
\caption{Average cost (Avg Cost) and task performance (Acc \%) achieved by BAMAS and baseline approaches on the GSM8K and MBPP datasets.}
\label{tab:main_comparison}
\resizebox{1\linewidth}{!}{%
\begin{tabular}{ll
                rr
                rr}
\toprule
\multirow{2}{*}{\textbf{Approach}} & \multirow{2}{*}{\textbf{Setting}}
& \multicolumn{2}{c}{\textbf{GSM8K}}
& \multicolumn{2}{c}{\textbf{MBPP}} \\
\cmidrule(lr){3-4} \cmidrule(lr){5-6}
& & \textbf{Avg Cost} & \textbf{Acc \%} 
  & \textbf{Avg Cost} & \textbf{Acc \%} \\
\midrule
AutoGen & DeepSeek-V3       & 1425.3  & \textbf{95.4} & 2661.3  & 80.8 \\
AutoGen & GPT-4.1 nano      &  475.3  & 89.7          &  666.1 & 71.4 \\
MetaGPT & DeepSeek-V3       & 3235.4 & 93.5          & 3735.1 & \textbf{82.2} \\
MetaGPT & GPT-4.1 nano      & 1012.3  & 86.8          & 1115.4 & 71.8 \\
ChatDev & DeepSeek-V3       & 2733.1 & 95.0          & 3635.1 & 81.2 \\
ChatDev & GPT-4.1 nano      &  800.5 & 90.1          & 1020.1 & 70.0 \\
\midrule
\multirow{5}{*}{\textbf{BAMAS}} 
        & Budget 500        &  222.4  & 87.9          &  153.7 & 73.8 \\
        & Budget 875        &  421.6 & 92.4          &  316.0 & 80.4 \\
        & Budget 1,250       &  447.0 & 94.9          &  529.2 & \textbf{82.6} \\
        & Budget 1,625       &  542.9  & \textbf{95.3} &  630.5 & 82.2 \\
        & Budget 2,000       &  542.6  & 95.1          &  811.0 & 82.2 \\
\bottomrule
\end{tabular}%
}
\end{table}

\begin{table}[t]
\scriptsize
\centering
\caption{Average cost (Avg Cost) and task performance (Acc \%) achieved by BAMAS and baseline approaches on the MATH dataset.}
\label{tab:math_comparison}
\begin{tabular}{llrrr}
\toprule
\textbf{Approach} & \textbf{Setting} & \textbf{Avg Cost} & \textbf{Acc \%} \\
\midrule
AutoGen & GPT-4.1 nano   &  797.2 & 77.6 \\
MetaGPT & GPT-4.1 nano   & 1380.0 & 67.1 \\
ChatDev & GPT-4.1 nano   & 1338.7 & 75.5 \\
\midrule
\multirow{3}{*}{\textbf{BAMAS}} 
        & Budget 1,000    &  339.4 & 72.1 \\
        & Budget 2,000    &  646.0 & 81.2 \\
        & Budget 3,000    &  870.7 & 82.7 \\
\bottomrule
\end{tabular}
\end{table}

\subsection{Configuration of BAMAS}\label{configurebamas}
BAMAS operates under a cost budget, with the range selected to cover meaningful operational regimes. The minimum budget is defined as the estimated cost of executing a viable workflow with two LLMs, which is 500 for GSM8K and MBPP, and 1,000 for MATH.
To examine BAMAS’s performance under different constraints, we set five budget levels for GSM8K and MBPP: 500, 875, 1,250, 1,650, and 2,000. For MATH, due to its higher execution cost, we use only three budget levels, namely 1,000, 2,000, and 3,000, to control expenses.

Additional hyperparameter configurations and experimental details are provided in Appendix~B.

\begin{figure*}
    \centering
    \includegraphics[width=1\linewidth]{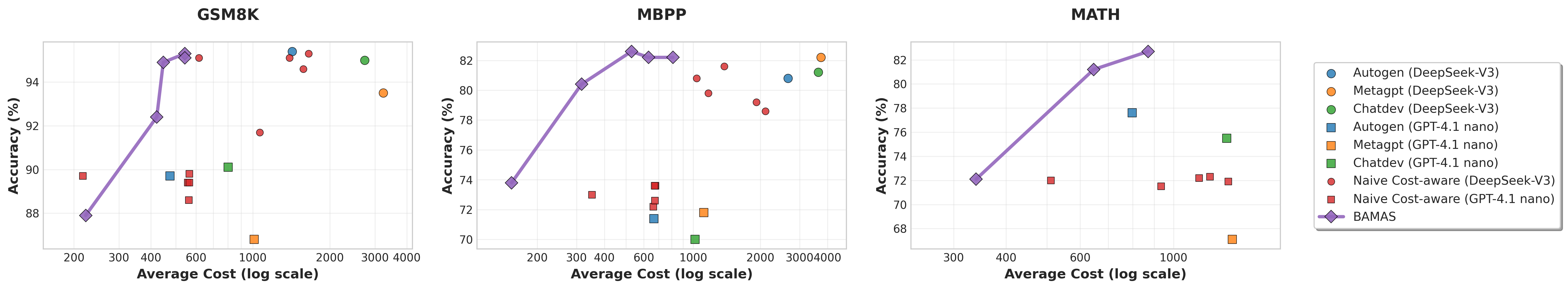}
    \caption{Average cost and accuracy achieved by BAMAS and baseline approaches across datasets.}
    \label{fig:enter-label}
\end{figure*}

\section{Results}

\subsection{RQ1: Cost-Performance Trade-Off}
This RQ evaluates how effectively BAMAS balances cost and task performance, i.e., the cost-performance trade-off, in comparison with existing agent construction approaches.

\paragraph{Cost and performance analysis.} Tables \ref{tab:main_comparison} and \ref{tab:math_comparison} present the average cost (Avg Cost) and task success rate (Acc \%) achieved by BAMAS and each baseline approach across the three datasets. As described earlier, each baseline approach is implemented in two versions, each using a different type of LLM, while BAMAS is implemented under varying budget constraints. For clarity, we also visualize the cost–performance results in a two-dimensional plot, as presented in Figure \ref{fig:enter-label}.

As shown in Figure \ref{fig:enter-label}, BAMAS exhibits an overall cost–performance trade-off. Across all three datasets, increasing the cost consistently leads to higher accuracy. Unlike existing approaches, BAMAS is tunable, allowing practitioners to adjust it to their budget constraints. In real-world applications, this tunability means that higher accuracy can be achieved simply by allocating a larger budget.

More importantly, BAMAS consistently demonstrates a superior cost–performance trade-off compared to existing approaches. As shown in Tables \ref{tab:main_comparison} and \ref{tab:math_comparison}, on GSM8K with a budget of 1,625, BAMAS achieves 95.3\% accuracy, nearly matching AutoGen with DeepSeek-V3 (95.4\%), while incurring an average cost of only 542.9 compared to 1,425.3, representing a 62\% cost reduction. Similarly, on the MBPP dataset, BAMAS reaches 82.6\% accuracy, comparable to the state-of-the-art (82.2\%), but at a cost of 529.2 versus 3,735.1, an 86\% reduction. On the MATH dataset, with a budget of 2,000, BAMAS surpasses existing approaches in accuracy (81.2\% vs. 77.6\%) while maintaining a lower cost (646.0 vs. 797.2). These results demonstrate that BAMAS can match or exceed state-of-the-art accuracy while using substantially fewer resources.

\begin{table}[t]
\centering
\footnotesize
\caption{Out-of-budget task counts for BAMAS across datasets and budget constraints. For example, ``1 / 500'' for MBPP under a budget of 500 means that, with this budget constraint, BAMAS exceeds the budget on only 1 out of 500 tasks in the MBPP dataset.}
\label{tab:budget_adherence}
\small
\begin{tabular}{r|r|r|r}
\toprule
\textbf{Budget} & \textbf{GSM8K} & \textbf{MBPP} & \textbf{MATH} \\
\midrule
500  & 0 / 1,319 & 1 / 500 & - \\
875  & 0 / 1,319 & 3 / 500 & - \\
1,000 & -        & -       & 11 / 1,000 \\
1,250 & 0 / 1,319 & 5 / 500 & - \\
1,625 & 0 / 1,319 & 3 / 500 & - \\
2,000 & 0 / 1,319 & 3 / 500 & 30 / 1,000 \\
3,000 & -        & -       & 0 / 1,000 \\
\bottomrule
\end{tabular}%
\end{table}

\paragraph{Budget adherence of BAMAS.} Having shown that BAMAS achieves a better cost–performance trade-off than existing approaches, we now examine its cost efficiency in more detail by assessing whether it adheres to budget constraints. Because the exact cost of executing a multi-agent system is non-deterministic, it is inevitable that any construction approach may occasionally produce systems whose execution cost exceeds the specified budget.

Table \ref{tab:budget_adherence} reports the out-of-budget (OOB) task counts for BAMAS across different datasets and budget constraints. Certain cells are left blank because these datasets are evaluated under different budget configurations, as specified in the ``Configuration of BAMAS'' section.
On the 1,319 tasks in the GSM8K dataset, BAMAS incurs no OOB tasks. For the MBPP dataset, the number of OOB tasks ranges from 1 to 5 across the five budget settings. For the MATH dataset, the highest OOB count is 30 under a budget of 2,000, still only 3\% of all tasks. Overall, OOB occurrences in BAMAS are rare.

\subsection{RQ2: Component Analysis}
\begin{table}[t]
\centering
\caption{Average cost (Avg Cost) and task performance (Acc \%) achieved by BAMAS and Naive-CostAware on the GSM8K and MBPP datasets.}
\label{tab:ablation}
\resizebox{1\linewidth}{!}{%
\begin{tabular}{ll
                rr
                rr}
\toprule
\multirow{2}{*}{\textbf{Approach}} & \multirow{2}{*}{\textbf{Setting}}
& \multicolumn{2}{c}{\textbf{GSM8K}}
& \multicolumn{2}{c}{\textbf{MBPP}} \\
\cmidrule(lr){3-4} \cmidrule(lr){5-6}
& & \textbf{Avg Cost} & \textbf{Acc (\%)} 
  & \textbf{Avg Cost} & \textbf{Acc (\%)} \\
\midrule
\multirow{5}{*}{Naive-CostAware} 
  & L1\&DeepSeek-V3 &  615.8 & 95.1 & 1037.1 & 80.8 \\
  & L2\&DeepSeek-V3 & 1063.5 & 91.7 & 1379.1 &  81.6 \\
  & L3\&DeepSeek-V3 & 1389.9 & 95.1 & 1770.1 & 79.8 \\
  & L4\&DeepSeek-V3 & 1574.8 & 94.6 & 1919.3 &  79.2 \\
  & L5\&DeepSeek-V3 & 1650.8 & 95.3 & 2106.7 &  78.6 \\
\midrule
\multirow{5}{*}{Naive-CostAware} 
  & L1\&GPT-4.1 nano &  216.7 & 89.7 &  352.1 & 73.0 \\
  & L2\&GPT-4.1 nano &  557.0 & 89.4 &  662.4 & 72.2 \\
  & L3\&GPT-4.1 nano &  562.4 & 88.6 &  673.4 &  72.6 \\
  & L4\&GPT-4.1 nano &  564.7 & 89.4 &  677.9 &  73.6 \\
  & L5\&GPT-4.1 nano &  565.2 & 89.8 &  671.8 &  73.6 \\
\midrule
\multirow{5}{*}{\textbf{BAMAS}} 
  & Budget 500  &  222.4 &  87.9 &  153.7 &  73.8 \\
  & Budget 875  &  421.6 & 92.4 &  316.0 &  80.4 \\
  & Budget 1,250 &  447.0 & 94.9 &  529.2 & 82.6 \\
  & Budget 1,625 &  542.9 & 95.3 &  630.5 &  82.2 \\
  & Budget 2,000 &  542.6 & 95.1 &  811.0 &  82.2 \\
\bottomrule
\end{tabular}%
}
\end{table}

This RQ evaluates the importance of LLM provisioning and topology selection in BAMAS by comparing it with Naive-CostAware, a baseline that greedily selects agent configurations according to five predefined resource levels. This baseline represents a simple, brute-force approach to cost management, operating in discrete levels ($L_i$), where each level $i$ uses a fixed set of $i+1$ identical agents and a static linear collaboration topology. In contrast, BAMAS decouples resource provisioning from strategy: its budget-constrained LLM provisioning determines an optimal LLM composition, while its agent collaboration topology selection chooses the most suitable topology.

Table \ref{tab:ablation} reports the average cost and task performance of BAMAS and Naive-CostAware. Due to space limits, we show results only for GSM8K and MBPP, though similar trends can be observed on the MATH dataset in Figure~\ref{fig:enter-label}.

We find that BAMAS achieves a superior cost–performance trade-off compared to Naive-CostAware, highlighting the necessity of both LLM provisioning and topology selection. Specifically, on the GSM8K dataset, BAMAS reaches the peak accuracy of 95.3\%, matching the peak accuracy of Naive-CostAware (L5\&DeepSeek-V3), but at a substantially lower average cost (542.9 vs. 1,650.8). Similarly, on the MBPP dataset, BAMAS attains higher peak accuracy (82.6\% vs. 81.6\%) while also incurring lower cost (529.2 vs. 1,379.1). These results demonstrate the effectiveness of BAMAS’s joint optimization of LLM provisioning and topology selection.

\subsection{RQ3: Topology Selection}
\begin{figure}[t]
    \centering
    \includegraphics[width=1\linewidth]{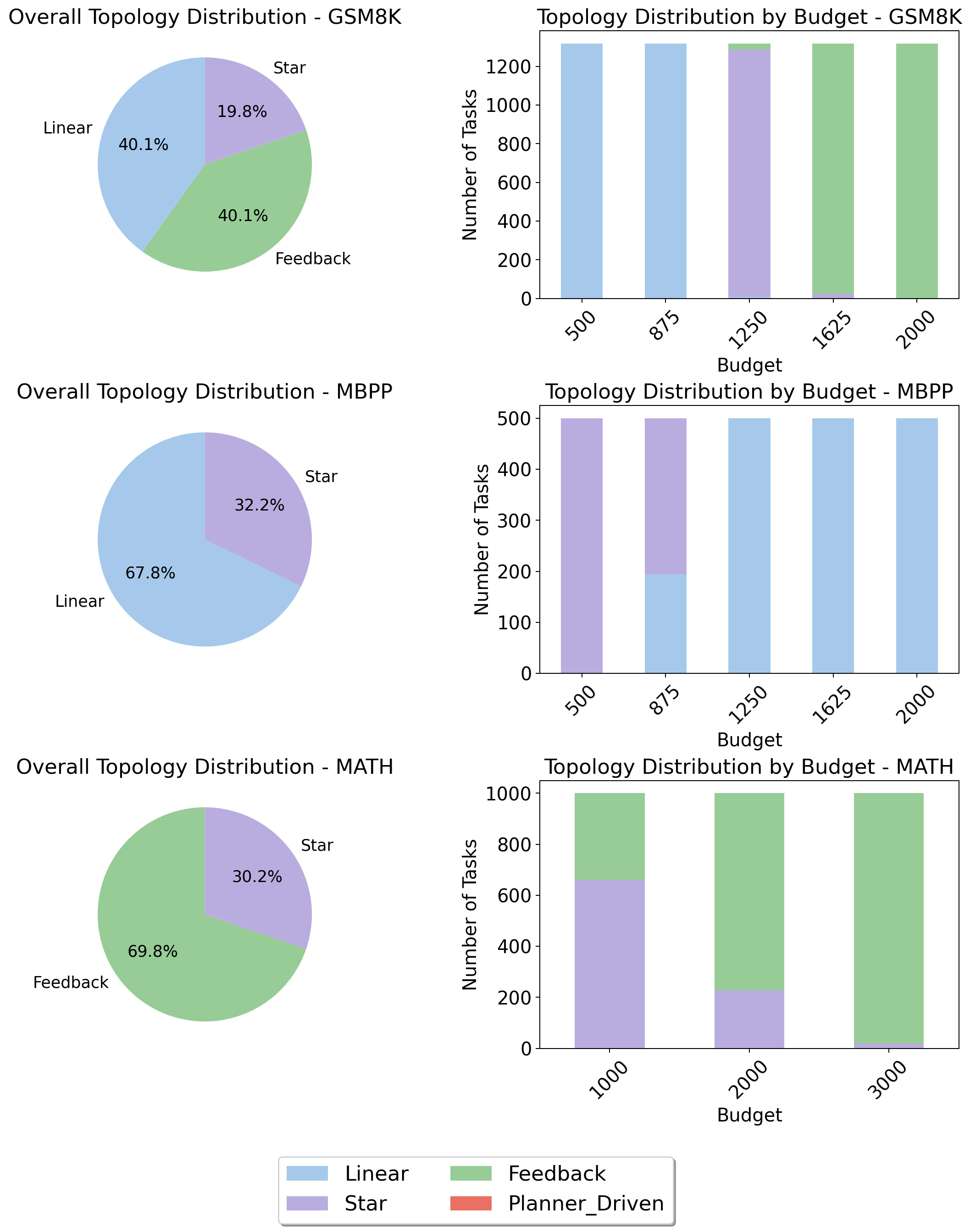}
    \caption{Collaboration topology distributions across datasets and budgets.}
    \label{fig:policy_choice}
\end{figure}

This RQ investigates the distribution of agent collaboration topologies to evaluate whether BAMAS can select diverse topologies across tasks and budget constraints. Figure~\ref{fig:policy_choice} presents the results: the left part shows the topology distributions across datasets, while the right part depicts, for each dataset, the distributions under varying budget constraints. From the figure, we make the following observations.

\paragraph{Task-specific topology selection.} BAMAS does not rely on a one-size-fits-all topology; rather, it adapts its strategy to the task domain. For mathematical reasoning tasks (i.e., tasks in the GSM8K and MATH datasets), which benefit from iterative refinement, the policy overwhelmingly converges on the \texttt{Feedback} topology. Specifically, BAMAS selects the \texttt{Feedback} topology for 40.1\% of tasks in GSM8K and 69.8\% of tasks in MATH. The higher proportion in MATH is expected, as it contains more challenging advanced problems compared to GSM8K. In contrast, for code generation tasks (i.e., the MBPP dataset), BAMAS predominantly favors the \texttt{Linear} topology, which aligns with a progressive refinement or multi-step generation process~\cite{seagentsurvey,agentfixagent}, making it a more effective and reliable strategy for such tasks.

\paragraph{Risk-averse and budget-dependent topology selection.} The topology selection policy of BAMAS is risk-averse under low budgets. Specifically, across all datasets, it heavily favors \texttt{Linear} and \texttt{Star} topologies when the budget is tight. This behavior is reasonable: these simpler topologies do not involve a critic agent, inherently reducing complexity and the likelihood of exceeding the budget. As the budget increases, allowing more flexibility, BAMAS becomes more willing to select complex topologies such as \texttt{Feedback}, which, while potentially more powerful, also carry higher baseline costs and greater risk of budget overrun. This demonstrates that the policy effectively adapts its strategy according to the available budget, balancing potential gains against financial risk.

\paragraph{Topology selection diversity.} Across all datasets and budgets, the \texttt{Planner-Driven} topology is never selected, a crucial insight into BAMAS’ topology selection policy. Although this topology offers the most flexibility, the reinforcement learning component learns that its high cost and instability (since a poor plan can derail the entire process) make it a suboptimal choice. Instead, the policy achieves higher and more reliable final rewards by relying on other, more structured patterns. Furthermore, we observe that the diversity of topology selection correlates with dataset characteristics. For the GSM8K dataset, topology choices for a given budget are highly uniform. In contrast, for the MATH and MBPP datasets, which contain more diverse problem types, the policy exhibits a more mixed distribution of topologies at certain budget levels. These findings suggest that BAMAS has learned a nuanced selection strategy, sensitive to the specific characteristics of problem instances.

%% file: 5_conclusion.tex
\section{Conclusion}
This paper presents BAMAS, a budget-aware approach for constructing multi-agent systems. BAMAS (1) uses Integer Linear Programming to allocate LLMs within budget constraints, (2) applies offline reinforcement learning to choose an effective collaboration topology, and (3) instantiates the system with the selected LLMs and topology.
The evaluation results show that BAMAS achieves a state-of-the-art cost–performance trade-off compared to existing agent construction approaches, attaining comparable or superior accuracy with substantially lower cost. 

%% file: appendix.tex
\section{Appendix A: Training of BAMAS's Topology Selection Policy}
This section describes the training procedure of BAMAS’s topology selection policy. The full procedure is presented in Algorithm \ref{alg:training}, and all notations used are summarized in Table~\ref{tab:notation}.

The topology selection policy (referred to as Topo-Selection in Algorithm \ref{alg:training}) determines \textit{how} provisioned LLM agents should collaborate for a given task and budget. Rather than hard-coding a single workflow, we define a discrete action space over commonly used collaboration patterns: Linear, Star, Feedback, and Planner-Driven topologies.

Conditioned on a compact representation of the task and the user-specified budget, the policy selects a topology that balances performance and cost. We train this selector using an offline reinforcement learning objective to avoid the latency and expense of online trial-and-error with LLM calls. Specifically, the learning objective combines task success with strong penalties for cost overruns and modest rewards for remaining under budget. Optimization is performed via a reward-weighted policy-gradient update with entropy regularization (REINFORCE) \citep{williams1992reinforce}.

\begin{table}[t]
\centering
\caption{Notations used in Algorithm~\ref{alg:training}.}
\label{tab:notation}
\renewcommand{\arraystretch}{1.1}
\resizebox{\linewidth}{!}{%
\begin{tabular}{llp{0.6\linewidth}}
\toprule
\textbf{Symbol} & \textbf{Type} & \textbf{Meaning} \\
\midrule
$\mathcal{D}$ & dataset & Offline experience dataset of tuples $\{(T_j,B_j,t_j,\text{outcome}_j)\}$. \\
$T_j$ & text & Task description for example $j$. \\
$B_j$ & scalar & Budget value associated with example $j$ (per-run budget). \\
$t_j$ & categorical & Action/topology index chosen for example $j$. \\
$\text{outcome}_j$ & structured & Execution result signals for example $j$ (e.g., correctness, token usage, etc.). \\
$R_{\text{final},j}$ & real & Final scalar reward for example $j$, computed by $\text{CalculateReward}(\text{outcome}_j,B_j)$. \\
$s_j$ & vector & State built from the embedding of $T_j$ and the scalar $B_j$. \\
$\pi_{\theta}(w\mid s)$ & policy & Topo-Selection policy: probability of choosing topology $t$ in state $s$. \\
$\theta,\ \theta^*$ & parameters & Current and best (selected) parameters of the Topo-Selection policy. \\
$\alpha$ & scalar & Learning rate. \\
$\beta$ & scalar & Entropy regularization coefficient. \\
$M$ & integer & Batch size. \\
$\text{logits}_j$ & vector & Pre-softmax scores over topology choices for example $j$. \\
$L^{\text{PG}}$ & real & Policy-gradient loss: $-\frac{1}{M}\sum_{j=1}^{M}\log\pi_{\theta}(t_j\mid s_j)\,R_{\text{final},j}$. \\
$H$ & real & Entropy of the categorical distribution induced by $\text{logits}_j$. \\
$\mathcal{L}_{\text{total}}$ & real & Total loss: $L^{\text{PG}} - \beta\, H$. \\
$R_{\text{current}}$ & real & Evaluation reward on the full dataset $\mathcal{D}$ for the current $\pi_\theta$. \\
$R_{\text{best}}$ & real & Best-so-far evaluation reward used for model selection. \\
\texttt{GetLogProbAndLogits} & proc & Returns $\log \pi_{\theta}(t_j\mid s_j)$ and $\text{logits}_j$. \\
\texttt{Entropy} & proc & Computes entropy of a categorical distribution parameterized by logits. \\
\texttt{EvaluateOnFullDataset} & proc & Evaluates $\pi_\theta$ on $\mathcal{D}$ for model selection. \\
\texttt{ILP} & model & Integer (Mixed-Integer) Linear Program used for budget-constrained LLM provisioning. \\
\texttt{Budget} & scalar & User-specified maximum allowed cost per run. \\
\texttt{Avg Cost} & real & Average cost per run (in units consistent with Section~\ref{sec:appendix_hyperparams}). \\
\texttt{OOB} & count & Out-of-Budget run count (number of runs exceeding the budget). \\
\texttt{max\_tokens} & integer & Maximum tokens an agent can generate per turn (dataset-specific). \\
\texttt{Planner/Executor/Critic} & roles & Agent roles used in collaboration topologies. \\
\bottomrule
\end{tabular}%
}
\end{table}

\begin{algorithm}[tb]
\caption{Training Algorithm of BAMAS's Topology Selection Policy}
\label{alg:training}
\textbf{Input}: Offline experience dataset $\mathcal{D}$, initial Topo-Selection policy $\pi_\theta$, optimizer \\
\textbf{Parameter}: Learning rate $\alpha$, entropy coefficient $\beta$, batch size $M$ \\
\textbf{Output}: Optimized Topo-Selection policy parameters $\theta^*$
\begin{algorithmic}[1] 
\STATE Initialize Topo-Selection policy $\pi_\theta$ with random weights $\theta$
\STATE Initialize $\theta^* \leftarrow \theta$, $R_{\text{best}} \leftarrow -\infty$
\FOR{each epoch = 1, 2, ...}
    \FOR{each batch $\{(T_j, B_j, t_j, \text{outcome}_j)\}_{j=1}^{M}$ from $\mathcal{D}$}
        \STATE $R_{\text{final}, j} \leftarrow \text{CalculateReward}(\text{outcome}_j, B_j)$ for $j=1, \dots, M$
        \STATE Create state $s_j$ from embedding of $T_j$ and scalar $B_j$
        \STATE $\log \pi_{\theta}(t_j|s_j), \text{logits}_j \leftarrow \text{GetLogProbAndLogits}(\pi_\theta, s_j, t_j)$
        \STATE $L^{\text{PG}} \leftarrow - \frac{1}{M} \sum_{j=1}^{M} \log \pi_{\theta}(t_j|s_j) \cdot R_{\text{final}, j}$
        \STATE $H \leftarrow \text{Entropy}(\text{Categorical}(\text{logits}_j))$
        \STATE $\mathcal{L}_{\text{total}} \leftarrow L^{\text{PG}} - \beta \cdot H$
        \STATE optimizer.zero\_grad()
        \STATE $\mathcal{L}_{\text{total}}.backward()$
        \STATE optimizer.step()
    \ENDFOR
    \STATE // Evaluate the updated policy on the entire dataset $\mathcal{D}$ for model selection
    \STATE $R_{\text{current}} \leftarrow \text{EvaluateOnFullDataset}(\pi_\theta, \mathcal{D})$
    \IF{$R_{\text{current}} > R_{\text{best}}$}
        \STATE $R_{\text{best}} \leftarrow $ $R_{\text{current}}$
        \STATE $\theta^* \leftarrow \theta$ \quad // Save the best policy parameters
    \ENDIF
\ENDFOR
\STATE \textbf{return} $\theta^*$
\end{algorithmic}
\end{algorithm}

\section{Appendix B: Hyperparameter Settings and Implementation Details}
\label{sec:appendix_hyperparams}
This section describes the hyperparameter settings and implementation details of our experiments.

\subsection{LLM Provisioning}
The ILP solver is configured to prioritize performance under a budget constraint. Our formulation follows standard integer (mixed-integer) programming practice \cite{ilp,milp,constraint_optimization}.

\subsubsection{ILP Solver Settings}
\begin{itemize}
    \item \textbf{Solver Backend:} We use the open-source \texttt{PuLP} library with the CBC (COIN-OR Branch-and-Cut) solver \cite{mitchell2011pulp,cbc_github}.
    \item \textbf{Execution Mode:} The solver is run in silent mode (\texttt{msg=0}) to suppress verbose output.
    \item \textbf{Solver Parameters:} We use the default settings for the time limit, MIP (Mixed-Integer Programming) gap, and thread count.
\end{itemize}

\subsubsection{Cost Estimation Parameters}
To estimate the operational cost for the ILP, we use the following parameters:
\begin{itemize}
    \item \textbf{Average Prompt Tokens:} 500 for GSM8K/MATH and MBPP.
    \item \textbf{Maximum Expandable Instances:} The maximum number of lower-tier LLMs that can be provisioned in place of a single higher-tier one is set to 5.
\end{itemize}

This budgeting setup aligns with recent work on cost-aware routing/cascades for LLMs \cite{chen2023frugalgpt,zhang2024treacle,dekoninck2025cascade}.

\subsection{Topo-Selection Policy Network}
The architecture and training parameters of our Topo-Selection policy  $\pi_\theta$ are detailed below.

\subsubsection{Policy Network Architecture}
The policy network processes a task embedding and the budget to produce a probability distribution over the available collaboration topologies.
\begin{itemize}
    \item \textbf{Task Encoder:} We use the \texttt{`all-MiniLM-L6-v2'} model to generate a 384-dimensional embedding for the input task description \cite{reimers2019sentencebert,wang2020minilm}.
    \item \textbf{Input Layers:}
        \begin{itemize}
            \item An \textit{Embedding Layer} projects the 384-dimensional task embedding to a 64-dimensional feature vector.
            \item A \textit{Budget Layer} projects the scalar budget value up to a 64-dimensional feature vector.
        \end{itemize}
    \item \textbf{Core Network:} The two 64-dimensional feature vectors are concatenated and fed into a Multi-Layer Perceptron (MLP) with one hidden layer of 128 neurons (\texttt{128 -> 128}).
    \item \textbf{Output Layer:} A final linear layer maps the 128-dimensional hidden state to the number of available topology patterns.
    \item \textbf{Activation Function:} The ReLU activation function is used throughout the network, except for the final output layer.
\end{itemize}

\subsubsection{Training Hyperparameters}
The Topo-Selection policy is trained using an offline policy gradient approach \cite{williams1992reinforce,levine2020offlinereinforcementlearningtutorial}. Hyperparameters are shared across datasets, except for the learning rate, which is the primary parameter adjusted for each specific task. The detailed settings are listed in Table \ref{tab:training_hyperparams}. 

\begin{table}[t]
\centering
\caption{Topo-Selection policy network training hyperparameters for each dataset.}
\label{tab:training_hyperparams}
\begin{tabular}{lrrr}
\toprule
\textbf{Hyperparameter} & \textbf{GSM8K} & \textbf{MATH} & \textbf{MBPP} \\
\midrule
Learning Rate & 0.0003 & 0.0015 & 0.0015 \\
Optimizer & Adam & Adam & Adam \\
Batch Size & 20,000 & 20,000 & 20,000 \\
Training Epochs & 10 & 10 & 10 \\
Entropy Coefficient & 0.001 & 0.001 & 0.001 \\
\bottomrule
\end{tabular}
\end{table}

We use Adam for optimization \cite{kingma2015adam}.

\subsection{Agent Instantiation}
The instantiation engine orchestrates the agents based on the selected topology and provisioned LLMs.

\subsubsection{Execution Control}
\begin{itemize}
    \item \textbf{Replanning Limit:} For the Planner-Driven topology, the \texttt{PlannerAgent} is permitted a maximum of two replanning attempts.
    \item \textbf{Budget Enforcement:} If the cumulative cost exceeds the allocated budget at any point, the execution is immediately terminated.
\end{itemize}

\subsubsection{Agent Token Limits}
The maximum number of tokens (\texttt{max\_tokens}) an agent can generate in a single turn was configured per dataset to match task complexity, as shown in Table \ref{tab:agent_tokens}.

\begin{table}[t]
\centering
\caption{Maximum generation tokens of BAMAS for agents on each dataset.}
\label{tab:agent_tokens}
\begin{tabular}{lrrr}
\toprule
\textbf{Agent Role} & \textbf{GSM8K} & \textbf{MATH} & \textbf{MBPP} \\
\midrule
Planner Agent & 384 & 384 & 384 \\
Executor Agent & 384 & 1,024 & 384 \\
Critic Agent & 384 & 1,024 & 384 \\
\bottomrule
\end{tabular}
\end{table}

\subsection{Other Experimental Settings}

\subsubsection{LLM Decoding Strategy}
To ensure deterministic and reproducible outputs from the LLM agents, we employ a greedy decoding strategy.
\begin{itemize}
    \item \textbf{Temperature:} Set to \texttt{0.0} for all LLM calls.
    \item \textbf{Top-p / Top-k:} Not used, consistent with greedy decoding.
    \item \textbf{Providers:} Our LLM asset library includes models from DeepSeek (\texttt{deepseek-chat}) and OpenAI (\texttt{gpt-4.1-nano}) \cite{deepseek_r1}.
\end{itemize}

\subsubsection{Reproducibility and Random Seeds}
We control for stochasticity by setting a global random seed at the beginning of each script. This affects neural network initialization, data shuffling, and the policy's sampling process. The ILP solver and the LLM decoding steps are deterministic.
\begin{itemize}
    \item \textbf{GSM8K Seed:} 42
    \item \textbf{MATH Seed:} 42
    \item \textbf{MBPP Seed:} 123
\end{itemize}

%% file: aaai2026.bib
@article{zhou2024survey,
  title={A Survey on LLM-Based Multi-Agent Systems: Workflow, Infrastructure, and Challenges},
  author={Li, Xinyi and Wang, Sai and Zeng, Siqi and Wu, Yu and Yang, Yi},
  journal={Vicinagearth},
  volume={1},
  number={1},
  pages={9},
  year={2024},
  publisher={Springer}
}

@inproceedings{hong2023metagpt,
  author       = {Sirui Hong and
                  Mingchen Zhuge and
                  Jonathan Chen and
                  Xiawu Zheng and
                  Yuheng Cheng and
                  Jinlin Wang and
                  Ceyao Zhang and
                  Zili Wang and
                  Steven Ka Shing Yau and
                  Zijuan Lin and
                  Liyang Zhou and
                  Chenyu Ran and
                  Lingfeng Xiao and
                  Chenglin Wu and
                  J{\"{u}}rgen Schmidhuber},
  title        = {MetaGPT: Meta Programming for {A} Multi-Agent Collaborative Framework},
  booktitle    = {Proceedings of the Twelfth International Conference on Learning Representations,
                  {ICLR} 2024},
  year         = {2024}
}

@inproceedings{wu2023autogen,
  title={{AutoGen}: Enabling Next-Gen {LLM} Applications via Multi-Agent Conversation},
  author={Qingyun Wu and
                  Gagan Bansal and
                  Jieyu Zhang and
                  Yiran Wu and
                  Shaokun Zhang and
                  Erkang Zhu and
                  Beibin Li and
                  Li Jiang and
                  Xiaoyun Zhang and
                  Chi Wang},
  booktitle={Proceedings of Conference on Language Modeling, COLM 2024},
  year={2024}
}

@article{seagentsurvey,
  author       = {Junwei Liu and
                  Kaixin Wang and
                  Yixuan Chen and
                  Xin Peng and
                  Zhenpeng Chen and
                  Lingming Zhang and
                  Yiling Lou},
  title        = {Large Language Model-Based Agents for Software Engineering: {A} Survey},
  journal      = {ACM Transactions on Software Engineering and Methodology},
  year         = {2025}
}

@inproceedings{agentfixagent,
  author       = {Alfin Wijaya Rahardja and
                  Junwei Liu and
                  Weitong Chen and
                  Zhenpeng Chen and
                  Yiling Lou},
  title        = {Can Agents Fix Agent Issues?},
  booktitle    = {Advances in Neural Information Processing Systems: Annual Conference
                  on Neural Information Processing Systems 2025, NeurIPS 2025},
  year         = {2025}
}

@misc{deepseek_r1,
  title={{DeepSeek-R1}: Pushing the Limits of Reasoning in {LLMs}},
  author={DeepSeek AI},
  year={2024},
  howpublished={Technical Report}
}

@book{ilp,
  title={Integer Programming},
  author={Wolsey, Laurence A},
  year={2020},
  publisher={Wiley}
}

@article{milp,
  author       = {Juan Pablo Vielma},
  title        = {Mixed Integer Linear Programming Formulation Techniques},
  journal      = {{SIAM} Rev.},
  volume       = {57},
  number       = {1},
  pages        = {3--57},
  year         = {2015}
}

@article{constraint_optimization,
  title={Constraint Optimization and Machine Learning},
  author={Kotthoff, Lars},
  journal={Artificial Intelligence Review},
  volume={53},
  pages={3233--3257},
  year={2020},
  publisher={Springer}
}

@inproceedings{wang-etal-2024-rethinking-bounds,
    title = "Rethinking the Bounds of {LLM} Reasoning: Are Multi-Agent Discussions the Key?",
    author = "Wang, Qineng  and
      Wang, Zihao  and
      Su, Ying  and
      Tong, Hanghang  and
      Song, Yangqiu",
    booktitle = "Proceedings of the 62nd Annual Meeting of the Association for Computational Linguistics (Volume 1: Long Papers), ACL 2024",
    year = "2024",
    pages = "6106--6131"
}

@misc{cobbe2021trainingverifierssolvemath,
      title={Training Verifiers to Solve Math Word Problems}, 
      author={Karl Cobbe and Vineet Kosaraju and Mohammad Bavarian and Mark Chen and Heewoo Jun and Lukasz Kaiser and Matthias Plappert and Jerry Tworek and Jacob Hilton and Reiichiro Nakano and Christopher Hesse and John Schulman},
      year={2021},
      eprint={2110.14168},
      archivePrefix={arXiv},
      primaryClass={cs.LG},
      url={https://arxiv.org/abs/2110.14168}, 
}

@misc{austin2021programsynthesislargelanguage,
      title={Program Synthesis with Large Language Models}, 
      author={Jacob Austin and Augustus Odena and Maxwell Nye and Maarten Bosma and Henryk Michalewski and David Dohan and Ellen Jiang and Carrie Cai and Michael Terry and Quoc Le and Charles Sutton},
      year={2021},
      eprint={2108.07732},
      archivePrefix={arXiv},
      primaryClass={cs.PL},
      url={https://arxiv.org/abs/2108.07732}, 
}

@inproceedings{hendrycks2021measuringmathematicalproblemsolving,
  author       = {Dan Hendrycks and
                  Collin Burns and
                  Saurav Kadavath and
                  Akul Arora and
                  Steven Basart and
                  Eric Tang and
                  Dawn Song and
                  Jacob Steinhardt},
  title        = {Measuring Mathematical Problem Solving With the {MATH} Dataset},
  booktitle    = {Proceedings of the Neural Information Processing Systems Track on
                  Datasets and Benchmarks, NeurIPS Datasets and Benchmarks 2021},
  year         = {2021}
}

@misc{levine2020offlinereinforcementlearningtutorial,
      title={Offline Reinforcement Learning: Tutorial, Review, and Perspectives on Open Problems}, 
      author={Sergey Levine and Aviral Kumar and George Tucker and Justin Fu},
      year={2020},
      eprint={2005.01643},
      archivePrefix={arXiv},
      primaryClass={cs.LG},
      url={https://arxiv.org/abs/2005.01643}, 
}

@inproceedings{qian2024chatdevcommunicativeagentssoftware,
  author       = {Chen Qian and
                  Wei Liu and
                  Hongzhang Liu and
                  Nuo Chen and
                  Yufan Dang and
                  Jiahao Li and
                  Cheng Yang and
                  Weize Chen and
                  Yusheng Su and
                  Xin Cong and
                  Juyuan Xu and
                  Dahai Li and
                  Zhiyuan Liu and
                  Maosong Sun},
  title        = {ChatDev: Communicative Agents for Software Development},
  booktitle    = {Proceedings of the 62nd Annual Meeting of the Association for Computational
                  Linguistics (Volume 1: Long Papers), {ACL} 2024},
  pages        = {15174--15186},
  year         = {2024}
}

@misc{chiang2024chatbotarenaopenplatform,
      title={Chatbot Arena: An Open Platform for Evaluating LLMs by Human Preference}, 
      author={Wei-Lin Chiang and Lianmin Zheng and Ying Sheng and Anastasios Nikolas Angelopoulos and Tianle Li and Dacheng Li and Hao Zhang and Banghua Zhu and Michael Jordan and Joseph E. Gonzalez and Ion Stoica},
      year={2024},
      eprint={2403.04132},
      archivePrefix={arXiv},
      primaryClass={cs.AI},
      url={https://arxiv.org/abs/2403.04132}, 
}

@article{williams1992reinforce,
  title   = {Simple Statistical Gradient-Following Algorithms for Connectionist Reinforcement Learning},
  author  = {Williams, Ronald J.},
  journal = {Machine Learning},
  volume  = {8},
  number  = {3-4},
  pages   = {229--256},
  year    = {1992},
  doi     = {10.1007/BF00992696},
  url     = {https://link.springer.com/article/10.1007/BF00992696}
}

@inproceedings{thomas2016data,
  title     = {Data-Efficient Off-Policy Policy Evaluation for Reinforcement Learning},
  author    = {Thomas, Philip S. and Brunskill, Emma},
  booktitle = {Proceedings of the 33rd International Conference on Machine Learning, ICML 2016},
  pages     = {3158--3166},
  year      = {2016}
}

@inproceedings{xie2019mis,
  author       = {Tengyang Xie and
                  Yifei Ma and
                  Yu{-}Xiang Wang},
  title        = {Towards Optimal Off-Policy Evaluation for Reinforcement Learning with
                  Marginalized Importance Sampling},
  booktitle    = {Proceedings of Annual Conference
                  on Neural Information Processing Systems 2019, NeurIPS 2019},
  pages        = {9665--9675},
  year         = {2019}
}

@article{chen2023frugalgpt,
  author       = {Lingjiao Chen and Matei Zaharia and James Zou},
  title        = {FrugalGPT: How to Use Large Language Models While Reducing Cost and
                  Improving Performance},
  journal      = {Trans. Mach. Learn. Res.},
  volume       = {2024},
  year         = {2024},
  url          = {https://openreview.net/forum?id=cSimKw5p6R},
  bibsource    = {dblp computer science bibliography, https://dblp.org}
}

@inproceedings{zhang2024treacle,
  title     = {Efficient Contextual LLM Cascades through Budget-Constrained Policy Learning},
  author    = {Zhang, Xuechen and Huang, Zijian and Taga, Ege Onur and Joe-Wong, Carlee and Oymak, Samet and Chen, Jiasi},
  booktitle = {Proceedings of Annual Conference
                  on Neural Information Processing Systems 2024, NeurIPS 2024},
  year      = {2024}
}

@article{dekoninck2025cascade,
  title   = {A Unified Approach to Routing and Cascading for LLMs},
  author  = {Jasper Dekoninck and Maximilian Baader and Martin Vechev},
  journal = {arXiv preprint arXiv:2410.10347},
  year    = {2025},
  doi     = {10.48550/arXiv.2410.10347},
  url     = {https://arxiv.org/abs/2410.10347}
}

@article{li2025budgetguidance,
  title   = {Steering LLM Thinking with Budget Guidance},
  author  = {Li, Junyan and Zhao, Wenshuo and Zhang, Yang and Gan, Chuang},
  journal = {arXiv preprint arXiv:2506.13752},
  year    = {2025},
  url     = {https://arxiv.org/abs/2506.13752}
}

@article{zellinger2025earlyabstention,
  title   = {Cost-Saving LLM Cascades with Early Abstention},
  author  = {Zellinger, Michael J. and Liu, Rex and Thomson, Matt},
  journal = {arXiv preprint arXiv:2502.09054},
  year    = {2025},
  doi     = {10.48550/arXiv.2502.09054},
  url     = {https://arxiv.org/abs/2502.09054}
}

@article{arora2025efficientreasoning,
  title   = {Training Language Models to Reason Efficiently},
  author  = {Arora, Daman and Zanette, Andrea},
  journal = {arXiv preprint arXiv:2502.04463},
  year    = {2025},
  url     = {https://arxiv.org/abs/2502.04463}
}

@misc{li2025largelanguagemodelinference,
      title={Large Language Model Inference Acceleration: A Comprehensive Hardware Perspective}, 
      author={Jinhao Li and Jiaming Xu and Shan Huang and Yonghua Chen and Wen Li and Jun Liu and Yaoxiu Lian and Jiayi Pan and Li Ding and Hao Zhou and Yu Wang and Guohao Dai},
      year={2025},
      eprint={2410.04466},
      archivePrefix={arXiv},
      primaryClass={cs.AR},
      url={https://arxiv.org/abs/2410.04466}, 
}

@misc{chen2025surveyllmbasedmultiagentsystem,
      title={A Survey on LLM-Based Multi-Agent System: Recent Advances and New Frontiers in Application}, 
      author={Shuaihang Chen and Yuanxing Liu and Wei Han and Weinan Zhang and Ting Liu},
      year={2025},
      eprint={2412.17481},
      archivePrefix={arXiv},
      primaryClass={cs.CL},
      url={https://arxiv.org/abs/2412.17481}, 
}

@inproceedings{du2023mad,
  author       = {Yilun Du and
                  Shuang Li and
                  Antonio Torralba and
                  Joshua B. Tenenbaum and
                  Igor Mordatch},
  title        = {Improving Factuality and Reasoning in Language Models through Multiagent
                  Debate},
  booktitle    = {Proceedings of the Forty-first International Conference on Machine Learning, {ICML} 2024},
  year         = {2024}
}

@inproceedings{zheng2023judge,
  title     = {Judging LLM-as-a-Judge with MT-Bench and Chatbot Arena},
  author    = {Zheng, Lianmin and Chiang, Wei-Lin and Sheng, Ying and Zhuang, Siyuan and Wu, Zhanghao and Zhuang, Yonghao and Lin, Zi and Li, Zhuohan and Li, Dacheng and Xing, Eric P. and Zhang, Hao and Gonzalez, Joseph E. and Stoica, Ion},
  booktitle = {Proceedings of Annual Conference
                  on Neural Information Processing Systems 2023, NeurIPS 2023},
  year      = {2023}
}

@inproceedings{madaan2023selfrefine,
  author       = {Aman Madaan and
                  Niket Tandon and
                  Prakhar Gupta and
                  Skyler Hallinan and
                  Luyu Gao and
                  Sarah Wiegreffe and
                  Uri Alon and
                  Nouha Dziri and
                  Shrimai Prabhumoye and
                  Yiming Yang and
                  Shashank Gupta and
                  Bodhisattwa Prasad Majumder and
                  Katherine Hermann and
                  Sean Welleck and
                  Amir Yazdanbakhsh and
                  Peter Clark},
  title        = {Self-Refine: Iterative Refinement with Self-Feedback},
  booktitle    = {Proceedings of Annual Conference
                  on Neural Information Processing Systems 2023, NeurIPS 2023},
  year         = {2023}
}

@inproceedings{shinn2023reflexion,
  author       = {Noah Shinn and
                  Federico Cassano and
                  Ashwin Gopinath and
                  Karthik Narasimhan and
                  Shunyu Yao},
  title        = {Reflexion: Language Agents with Verbal Reinforcement Learning},
  booktitle    = {Proceedings of Annual Conference
                  on Neural Information Processing Systems 2023, NeurIPS 2023},
  year         = {2023}
}

@inproceedings{yao2022react,
  author       = {Shunyu Yao and
                  Jeffrey Zhao and
                  Dian Yu and
                  Nan Du and
                  Izhak Shafran and
                  Karthik R. Narasimhan and
                  Yuan Cao},
  title        = {ReAct: Synergizing Reasoning and Acting in Language Models},
  booktitle    = {Proceedings of the Eleventh International Conference on Learning Representations,
                  {ICLR} 2023},
  year         = {2023}
}

@inproceedings{yao2023treeofthoughts,
  author       = {Shunyu Yao and
                  Dian Yu and
                  Jeffrey Zhao and
                  Izhak Shafran and
                  Tom Griffiths and
                  Yuan Cao and
                  Karthik Narasimhan},
  title        = {Tree of Thoughts: Deliberate Problem Solving with Large Language Models},
  booktitle    = {Proceedings of Annual Conference
                  on Neural Information Processing Systems 2023, NeurIPS 2023},
  year         = {2023}
}

@inproceedings{reimers2019sentencebert,
  title        = {Sentence-BERT: Sentence Embeddings using Siamese BERT-Networks},
  author       = {Reimers, Nils and Gurevych, Iryna},
  booktitle    = {Proceedings of the 2019 Conference on Empirical Methods in Natural Language Processing and the 9th International Joint Conference on Natural Language Processing, EMNLP-IJCNLP 2019},
  pages        = {3982--3992},
  year         = {2019}
}

@inproceedings{wang2020minilm,
  title     = {{MiniLM}: Deep Self-Attention Distillation for Task-Agnostic Compression of Pre-Trained Transformers},
  author    = {Wang, Wenhui and Wei, Furu and Dong, Li and Bao, Hangbo and Yang, Nan and Zhou, Ming},
  booktitle = {Proceedings of Annual Conference
                  on Neural Information Processing Systems 2020, NeurIPS 2020},
  year      = {2020}
}

@inproceedings{kingma2015adam,
  title     = {Adam: A Method for Stochastic Optimization},
  author    = {Kingma, Diederik P. and Ba, Jimmy},
  booktitle = {Proceedings of the 3rd International Conference on Learning Representations, ICLR 2015},
  year      = {2015}
}

@misc{mitchell2011pulp,
  title        = {{PuLP}: A Linear Programming Toolkit for Python},
  author       = {Mitchell, Stuart and O'Sullivan, Michael and Dunning, Iain},
  year         = {2011},
  howpublished = {Optimization Online},
  url          = {https://optimization-online.org/2011/09/3178/}
}

@misc{cbc_github,
  title        = {{CBC}: COIN-OR Branch-and-Cut Mixed Integer Programming Solver},
  author       = {{COIN-OR Foundation}},
  year         = {2005--},
  howpublished = {GitHub Repository},
  url          = {https://github.com/coin-or/Cbc}
}
